\documentclass[aip,amsmath,amssymb,rsi,groupedaddress]{revtex4-1}
\usepackage{graphicx}
\usepackage{epsfig}
\usepackage{amsmath}
\usepackage{graphics}
\usepackage{epsfig}
\usepackage{yfonts}
\usepackage{amsmath}
\usepackage{amsfonts}
\usepackage{amssymb}
\usepackage{xcolor,cancel}

\begin{document}

\title{Non-adiabatic corrections to electric current in molecular junctions due to nuclear motion at the molecule-electrode interfaces}
\author{Vincent F. Kershaw and Daniel S. Kosov}
\address{College of Science and Engineering, James Cook University, Townsville, QLD, 4811, Australia 
}


\begin{abstract}
We present quantum electron transport theory that incorporates dynamical effects of motion of atoms on electrode-molecule interfaces in the calculations of the electric current. The theory is based on non-equilibrium Green's functions. We separate time scales in the Green's functions on fast relative time and slow central time. The derivative with respect to the central time serves as a small parameter in the theory. We solve the real-time Kadanoff-Baym equations for molecular Green's functions using Wigner representation and keep terms up to the second order with respect to the central time derivatives. Molecular Green's functions and consequently the electric current are expressed as functions of molecular junction coordinates as well as velocities and accelerations of molecule-electrode interface nuclei. We apply the theory to model a molecular system and study the effects of non-adiabatic  nuclear motion on molecular junction conductivity.
\end{abstract}

\maketitle

\section{Introduction}
Quantum transport of electrons through nanoscale molecular systems is an active field of research, which has made remarkable fundamental advances in recent years. This includes not only developing, after a decade of struggle, robust and reproducible experimental measurements but also obtaining the fundamental theoretical understanding of microscopic mechanisms of molecular quantum transport.\cite{thoss-evers-review,nichols2010}

Unfortunately, this scientific progress has not been transplanted in electronic devices for real world applications. Molecular electronics has for decades been touted as something to replace semiconductor electronics, but one major difficulty has dampened hopes.  Single-molecular junctions are sensitive to every microscopic static and dynamical detail of the electrode-molecule interface with the details not being possible to control. The thorough scientific understanding of molecular contacts is still required before the commercial potential of single-molecular technologies can be realised in electronics.

Not only is the interface geometry largely not known in a molecular junction but nuclear dynamics for the interfacial atoms play a critical role (owing to the gold-electron plasticity, significant voltage drop on molecule-metal interface creating a strong electric field and comparative weakness of the molecule-metal bonds).\cite{doi:10.1021/jacs.7b13694,C8CP00317C}
 The nuclear motion of molecule-electrode interfacial atoms can be considered as vibrational motion; standard theoretical techniques such as
 non-equilibrium Green's functions,\cite{caroli72,ryndyk06,dahnovsky:014104,galperin06,ryndyk07,hartle08,rabani14,hartle15,PhysRevB.75.205413}
 master equations \cite{PhysRevB.69.245302,fcblockade05,PhysRevB.83.115414,may02,PhysRevB.94.201407,segal15,kosov17-wtd,kosov17-nonren,dzhioev14,dzhioev15}
or scattering theory \cite{ness05,PhysRevB.70.125406,peskin07,doi:10.1063/1.3231604}  can, in principle, be applied to treat it (although the problem is technically harder for the theoretical treatment, since the vibrations are not localized in the central region\cite{peskin07,peskin17}).
With only a few recent exceptions,\cite{fuse,dzhioev11,bode12,catalysis12,galperin15,subotnik17-prl,kershaw17} all these standard theoretical approaches have to assume that the amplitudes of nuclear motions are small. Furthermore, they require that either electron-vibration coupling or interaction between the molecule and electrodes should be small in comparison with other energy scales in the system.

In this paper, we continue our development of a non-equilibrium Green's function based transport theory that takes into consideration non-adiabatic effects of nuclear motion. \cite{kershaw17} 
 The approach is based on the gradient expansion of the non-equilibrium Green's functions. The idea of using the gradient expansion to separate time and length scales in Green's functions goes back to the work of Kadanoff and Baym,\cite{kadanoff-baym} with the technique being perfected over years in the studies of non-equilibrium processes in nuclear 
and condensed matter physics.\cite{danielewicz84a,danielewicz84b,botermans1990,rammer86,rammer-book}
Recently, several studies\cite{bode12,fuse,catalysis12,PhysRevB.92.235440,galperin15,subotnik17-prl,kershaw17,subotnik18} used gradient expansions to treat dynamics of classical degrees of freedom as a slow varying disturbance in the electronic non-equilibrium Green's functions and this work follows the same philosophy. 
In our previous paper\cite{kershaw17} we computed the non-adiabatic correction to the electric current from non-adiabatic effects associated with nuclear motion in the central region. Here we extend the theory to include the interfacial nuclear dynamics from non-rigid molecule-electrode bonding.  Our approach is not based on the typical assumption that the amplitude of nuclear motion is either small or harmonic, nor is it required that the electron-nuclear coupling is small. The theory uses the velocity of nuclear motion as a small parameter and, consequently,  there are no restrictions on the scale of possible molecular conformational changes or strength of electron-vibrational interaction in our approach.

The outline of the paper is as follows. Section II contains the theory:  separable approximation for electrode self-energies,   solution of the real-time Kadanoff-Baym equations for molecular Green's functions using Wigner representation and the derivation of non-adiabatic formulae for  electric current. In section III we illustrate the proposed theory by the application to electron transport through a single resonant-level molecular junction with non-rigid molecule-electrode linkage. Section IV gives conclusions and  a summary of the main results.
We use atomic units in the derivations throughout the paper ($\hbar= |e|= m_e= 1$).

\section{Theory}
\subsection{Model Hamiltonian}
We consider a molecular junction: a single molecule connected to two macroscopic leads; the Hamiltonian for this system is given by
\begin{equation}
H=H_M + H_L + H_R + H_{LM} + H_{RM}.
\end{equation}
Here $H_M$ is the Hamiltonian for the molecule, $H_L$ is the Hamiltonian for the left lead and $H_R$ is the Hamiltonian for the right lead. The terms $H_{LM}$ and $H_{RM}$ describe the interactions between the molecule and the left and right leads, respectively. 
The molecule is modelled by a single molecular orbital with energy $\epsilon$ as
\begin{equation}
H_M=\epsilon  d^\dag d + V_N(q_L,q_R).
\end{equation}
Here $d^\dag$ and $d$ are fermionic creation and annihilation operators for a molecular electron.
Classical variables  $q_L$ and $q_R$  describe dynamical changes in the bond lengths between the molecule and the left and right leads, respectively; and  $V_N$ is the potential energy surface for these two variables.

The left and right leads of the molecular junction are modelled as macroscopic reservoirs of non-interacting electrons as
\begin{eqnarray}
H_L + H_R =  \sum_{k\alpha} \epsilon_{k\alpha} a^\dag_{k\alpha} a_{k\alpha},
\end{eqnarray}
where $a^\dagger_{k\alpha}$   creates  an electron in the single-particle state $k$  with energy $\epsilon_{k\alpha}$ of ($\alpha=L/R$) left/right leads  and $a_{k\alpha}$ is the corresponding electron  annihilation operator.   The lead-molecule couplings are described by the tunneling interaction
\begin{eqnarray}
H_{LM}(t) + H_{RM}(t)=  \sum_{k \alpha } (  v_\alpha(q_\alpha) a^\dag_{k\alpha} d +  v_\alpha^*(q_\alpha)d^\dag a_{k\alpha }),
\end{eqnarray}
where $v_{\alpha}(q_\alpha)$  are the tunnelling amplitudes between leads single-particle states and molecular orbital. The molecule-lead interaction is explicitly time-dependent due to the changes of the corresponding bond lengths from the equilibrium values $q_\alpha=x_\alpha -x_{0\alpha}$,  where $x_{0\alpha}$ is the equilibrium bond length. We assume that the tunnelling amplitudes have a linear dependence on $q_\alpha$, where it follows from this assumption that $v_{\alpha}(q_\alpha)$ takes the form
 \begin{equation}
 v_\alpha(q_\alpha) = (1+ \lambda_\alpha q_\alpha)  u_\alpha.
\label{intermsofx}
 \end{equation}
Here  $u_\alpha$ is the strength of the tunneling interaction and $\lambda_\alpha$ is the real parameter which describes modulation of the tunnelling amplitude due to the changes of the molecule-lead geometry. The assumption of the linear dependence of the tunneling amplitude on the nuclear coordinates is not critical for the derivation of the main equations; we can also carry out the similar derivations for completely arbitrary dependence of $v_\alpha$ on $q_\alpha$.

\subsection{Green's functions  and self-energies}

\subsubsection{Green's functions}
We define the exact (non-adiabatic, computed with a fully time-dependent Hamiltonian  along a given trajectory $\mathbf q(t)=(q_L(t),q_R(t))$) retarded, advanced and lesser Green's functions in a standard way as:\cite{haug-jauho}

\begin{equation}
{\cal G}^R(t,t') = -i \theta(t-t') \langle \{d (t), d^\dag (t')\} \rangle,
\end{equation}
\begin{equation}
{\cal G}^A(t,t') = \Big({\cal G}^R(t',t) \Big)^*
\end{equation}
and
\begin{equation}
{\cal G}^<(t,t') = i \langle d^\dag (t')d (t) \rangle.
\end{equation}

\subsubsection{Self-energies in time domain}
The influence of the electrodes on the molecular Green's function is taken into account via electrode self-energies. Left and right retarded self-energies are given by
\begin{equation}
{\Sigma}_{\alpha}^R(t,t') = -i \theta(t-t')  v^*_{\alpha} (t) v_{ \alpha} (t')  \sum_{k} e^{-i \epsilon_{k} (t-t')}.
\label{sigmaRt}
\end{equation}
Here $v_{k\alpha}(t)$ means $v_{k\alpha}\big(q_\alpha(t)\big)$. The advanced and retarded self-energies are related to each other  via Hermitian  conjugation:
\begin{equation}
{\Sigma}_{\alpha }^A(t,t') = \Big( {\Sigma}_{\alpha}^R(t',t) \Big)^*.
\label{sigmaAt}
\end{equation}
The lesser self-energy is defined as
\begin{equation}
{\Sigma}_{\alpha}^<(t,t') =i  v^*_{\alpha} (t) v_{ \alpha} (t') \sum_{k}   f_{\alpha}(\epsilon_k)  e^{-i \epsilon_{k} (t-t')},
\label{sigma<t}
\end{equation} 
where $f_\alpha$ is Fermi-Dirac occupation number for $\alpha=L,R$ electrodes. The total self-energies are the sum of contributions from the left and right electrodes
\begin{equation}
{\Sigma}^{R,A,<}(t,t') =  {\Sigma}^{R,A,<}_{L}(t,t')+ {\Sigma}_{R}^{R,A,<}(t,t').
\end{equation}

\subsubsection{Self-energies in Wigner representation and separable approximation}
We will solve the real-time Kadanoff-Baym equations using Wigner representation and to be able to do so we must first to convert electrode self-energies to the Wigner representation. Let us introduce central and relative times
\begin{equation}
T=\frac{1}{2} (t+t')
\end{equation}
and
\begin{equation}
\tau = t - t'
\end{equation}
for Green's functions ${\cal G}(t,t')$ and self-energies $\Sigma(t,t')$. The Wigner transformation is defined as the Fourier transformation with respect to relative time:
\begin{equation}
{ \widetilde \Sigma}(T, \omega) = \int^{+\infty}_{-\infty} d\tau e^{i \omega \tau} \Sigma ( T,\tau).
\end{equation}

For the calculation of the Wigner transformed  self-energies we propose a separable approximation  (separable functional form  with respect to central $T$ and relative $\tau$ times):\begin{equation}
{\Sigma}_{\alpha}^R(T,\tau) = -i \theta(\tau)  v^*_{\alpha}(T+\tau/2) v_{\alpha}(T-\tau/2)  \sum_{k}   e^{-i \epsilon_{k\alpha} \tau} \simeq -i \theta(\tau)  |v_{\alpha} (T)|^2  \sum_{k}    e^{-i \epsilon_{k\alpha} \tau}
\label{sigmaRt-appr}
\end{equation}
and
\begin{equation}
{\Sigma}_{\alpha}^<(T,\tau) =i  v^*_{\alpha}(T+\tau/2) v_{\alpha}(T-\tau/2)  \sum_{k}   f_\alpha(\epsilon_k) e^{-i \epsilon_k \tau} \simeq i  |v_{\alpha} (T)|^2   \sum_{k}   f_\alpha(\epsilon_k) e^{-i \epsilon_k \tau}.
\label{sigma<t-appr}
\end{equation} 
The separable approximation does not violate the standard relations between self-energies, for example $\Sigma_\alpha^A(T,\tau)= (\Sigma_\alpha^R(T,-\tau))^*$. We justify the use of a separable approximation  based on the following considerations. For the smooth lead's density of state  both sums $ \sum_{k} e^{-i \epsilon_{k\alpha} \tau}$ and $ \sum_{k}   f(\epsilon_k)  e^{-i \epsilon_{k\alpha} \tau}$ are peaked around $\tau=0$, thereby removing $\pm \tau/2$ time shifts from the tunnelling coupling amplitudes $v_\alpha$.

The use of the  separable approximation yields the following self-energies in Wigner space:
\begin{equation}
\widetilde \Sigma_\alpha^A(T,\omega)= \Lambda_\alpha(T,\omega) +\frac{i}{2} \Gamma_\alpha(T,\omega),
\end{equation}
\begin{equation}
\widetilde \Sigma_\alpha^R(T,\omega)= \Lambda_\alpha(T,\omega) - \frac{i}{2} \Gamma_\alpha(T,\omega)
\end{equation}
and
\begin{equation}
\widetilde  \Sigma_\alpha^<(T,\omega)= f_\alpha(\omega)\left( \Sigma^A(T,\omega) - \Sigma^R(T\omega)\right)= i f_\alpha(\omega) \Gamma_\alpha(T,\omega).
\end{equation}
Here
\begin{equation}
\Gamma_\alpha(T,\omega) =2 \pi |v_\alpha(T)|^2 \rho_\alpha(\omega),
\end{equation}
where representing $v_\alpha(T)$ in terms of (\ref{intermsofx}) gives
\begin{equation}
 \Gamma_\alpha(T,\omega) =2 \pi |(1+\lambda_\alpha q_\alpha) u_{\alpha}|^2 \rho_\alpha(\omega) = (1+\lambda_\alpha q_\alpha)^2 \gamma_\alpha(\omega).
\label{111}
\end{equation}
In the last equality above we have grouped together constants by defining the quantity 
\begin{equation}
\gamma_\alpha(\omega) =2 \pi |u_\alpha|^2 \rho_\alpha(\omega),
\label{112}
\end{equation}
which can be understood as the standard level-broadening function for a static molecular junction.

In what follows, we choose to work in the wide-band limit where $\rho_\alpha(\omega)$ is an energy independent constant (and hence $\gamma_\alpha(\omega)$ by (\ref{112})). In this limit, the self-energy components take the form:
\begin{equation}
\widetilde \Sigma_\alpha^A(T)=\frac{i}{2} \gamma_\alpha (1+\lambda_\alpha q_\alpha(T))^2 = \frac{i}{2}\Gamma_{\alpha}(T),
\label{advancedS}
\end{equation}
\begin{equation}
\widetilde \Sigma_\alpha^R(T)=  - \frac{i}{2} \gamma_\alpha (1+\lambda_\alpha q_\alpha(T))^2 =  - \frac{i}{2}\Gamma_{\alpha}(T)
\end{equation}
and
\begin{equation}
\widetilde  \Sigma_\alpha^<(T,\omega)= i f_\alpha(\omega)  \gamma_\alpha (1+\lambda_\alpha q_\alpha(T))^2 =  i f_\alpha(\omega) \Gamma_{\alpha}(T).
\label{lesserS}
\end{equation}
Notice that the retarded/advanced self-energies have lost their energy dependence on $\omega$ in the wide-band limit, they depend only on central time $T$. It is also important to highlight that the function $\Gamma_\alpha$ is also energy independent and takes the form 
\begin{equation}
\Gamma_{\alpha}(T) = \gamma_\alpha (1+\lambda_\alpha q_\alpha(T))^2.
\label{115}
\end{equation}

\subsection{Solution of real time Kadanoff-Baym equation via separation of time-scales}
We begin with the equation of motion for the non-adiabatic retarded Green's function (only this type of Green's functions will be later required for the electric current calculations):
\begin{eqnarray}
\Big( i\partial_t  - \epsilon \Big) {\cal G }^R(t,t') = 
 \delta(t-t') +  \int^{+\infty}_{-\infty} dt_1 \Sigma^R(t,t_{1}){\cal G }^R(t_{1},t').
  \label{eom-1}
\end{eqnarray}
The equation of motion  in  the Wigner representation becomes
\begin{eqnarray}
\Big( \omega +\frac{i}{2} \partial_T  - \epsilon \Big)   {\cal \widetilde G}^R(T,\omega)=I+ e^{\frac{1}{2i}(\partial^{\Sigma}_T \partial^{\cal G}_\omega- \partial^{\cal G}_T \partial^\Sigma_\omega)} \widetilde \Sigma^R(T,\omega) {\cal \widetilde  G}^R(T,\omega).
\end{eqnarray}
Here $\partial^\Sigma$ means the derivative acting on the self-energy only and $\partial^G$ denotes the derivative acting on the Green's function.
In the wide-band approximation $\widetilde \Sigma^R$  depends on the central time $T$  only and, consequently, the exponential operator acting on the retarded self-energy is simplified and we get
\begin{eqnarray}
\Big(\omega +\frac{i}{2} \partial_T  - \epsilon \Big)   {\cal \widetilde G}^{R}(T,\omega)=I+ e^{\frac{1}{2i}\partial^{\Sigma}_T \partial^{\cal G}_\omega} \widetilde \Sigma^R(T) {\cal \widetilde  G}^R(T,\omega).
\end{eqnarray}
We solve this Kadanoff-Baym equation  using the time derivative with respect to the central time as a small parameter.  It means we assume that the changes of the self-energies and the Green's functions are slow with respect to the central time and fast with respect to the relative time. The central time dependence is associated with slow nuclear dynamics (through the dependence of the self energy on classical variable $\mathbf q(t)$) and relative time oscillations are related to electronic time-scale (in our case the characteristic tunneling time for the electron to transport across the molecule).
The solution described below follows the general  ideas discussed in our previous paper.\cite{kershaw17}
Expanding the exponential operator up to the second order in the time derivatives we get a truncated equation of motion for the retarded Green's function
\begin{equation}
\Big( \omega +\frac{i}{2} \partial_T  - \epsilon \Big)   {\cal \widetilde G}^R=1+ \Big(\widetilde{\Sigma}^R + \frac{1}{2i}\partial_T \widetilde{\Sigma}^R \partial_\omega - \frac{1}{8}\partial^2_T \widetilde{\Sigma}^R \partial_\omega^2\Big) {\cal \widetilde  G}^R.
\label{eom-AR}
\end{equation}
Here we omit $T$ and $\omega$ variables from Green's functions and self energies for brevity.
We use the ansatz
\begin{equation}
\label{expansion}
\widetilde {\cal G}^R = \widetilde {\cal G}^R_{(0)}+ \widetilde{\cal G}^R_{(1)} + \widetilde{\cal G}^R_{(2)},
\end{equation}
when looking for the solution that contains the time derivatives up to the second order in the retarded Green's function. Here the term $ \widetilde {\cal G}^R_{(0)}$ depends on nuclear geometry only, $ \widetilde {\cal G}^R_{(1)}$ depends on nuclear geometry and is linearly proportional to the nuclear velocities and $ \widetilde {\cal G}^R_{(2)}$ which has dependencies on nuclear geometry, acceleration and is quadratic in velocities.

Substituting (\ref{expansion}) into  (\ref{eom-AR}) we obtain  a system of three equations based on order of the derivatives with respect to the central time: 
\begin{equation}
\Big( \omega  - \epsilon \Big)   \widetilde{\mathcal{G}}_{(0)}^R = 1 + \widetilde{\Sigma}^R \widetilde{\mathcal{G}}_{(0)}^R,
\end{equation}
\begin{equation}
\frac{i}{2} \partial_T  \widetilde{\mathcal{G}}_{(0)}^R + \Big( \omega  - \epsilon \Big)   \widetilde{\mathcal{G}}_{(1)}^R = \widetilde{\Sigma}^R \widetilde{\mathcal{G}}_{(1)}^R + \frac{1}{2i}\partial_T \widetilde{\Sigma}^R \partial_\omega \widetilde{\mathcal{G}}_{(0)}^R
\end{equation}
and 
\begin{equation}
\frac{i}{2} \partial_T  \widetilde{\mathcal{G}}_{(1)}^R + \Big( \omega  - \epsilon \Big)   \widetilde{\mathcal{G}}_{(2)}^R = \widetilde{\Sigma}^R \widetilde{\mathcal{G}}_{(2)}^R + \frac{1}{2i}\partial_T \widetilde{\Sigma}^R \partial_\omega \widetilde{\mathcal{G}}_{(1)}^R  - \frac{1}{8}\partial^2_T \widetilde{\Sigma}^R \partial^2_\omega \widetilde{\mathcal{G}}_{(0)}^R.
\end{equation}
The equation for the zeroth order Green's function is easily solved and gives
\begin{equation}
\widetilde {\cal G}^R_{(0)}  = \Big( \omega -\epsilon -\widetilde{\Sigma}^R \Big)^{-1} = G^R,
\end{equation}
which is the standard adiabatic retarded Green's function $G^R$. To solve for the first order correction we rearrange the respective equation in terms of $\widetilde{\mathcal{G}}_{(1)}^R$ to get
\begin{equation}
\widetilde {\cal G}^R_{(1)}  = - \frac{i}{2} G^R \partial_T  G^R  + \frac{1}{2i} G^R \partial_T \widetilde{\Sigma}^R \partial_\omega G^R.
\end{equation}
We note that 
\begin{equation}
\partial_\omega G^R = - \Big( G^R \Big)^2 
\end{equation}
and 
\begin{equation}
\partial_T G^R = \partial_T \widetilde{\Sigma}^R \Big( G^R \Big)^2,
\end{equation}
where these derivatives gives 
\begin{equation}
\widetilde {\cal G}^R_{(1)}  = 0.
\end{equation}
Therefore, the first order non-adiabatic correction to the retarded Green's function vanishes.
Now considering the second order correction, we rearrange for $\widetilde{\mathcal{G}}_{(2)}^R$ and make a substitution for $\widetilde{\mathcal{G}}_{(1)}^R$ to get
\begin{equation}
\widetilde{\mathcal{G}}_{(2)}^{R} =  - \frac{1}{8} G^{R} \partial^2_T \widetilde{\Sigma}^{R} \partial^2_\omega G^{R},
\end{equation}
which, after computing the double derivative of the adiabatic retarded Green's function, can be easily shown to produce
\begin{equation}
\widetilde{\mathcal{G}}_{(2)}^{R} =  - \frac{1}{4} \Big(G^{R} \Big)^{4} \partial^2_T \widetilde{\Sigma}^{R}.
\end{equation}
Here $\partial^2_T \widetilde{\Sigma}^{R}_\alpha$ is the second central time derivative of the retarded self-energy component and is to be given an explicit form later. 

\subsection{Formula for electric current}
We begin with the general expression for the electric current at time $t$ flowing from  $\alpha=L,R$ electrode to the molecule\cite{haug-jauho}
\begin{equation}
J_\alpha(t) = C_\alpha(t,t),
\end{equation}
where
\begin{equation}
C_\alpha(t,t') =2  \int_{-\infty}^{+\infty} dt_1 \text{Re} \Big\{ {\cal G}^<(t,t_1)\Sigma_\alpha^A(t_1,t')+ {\cal G}^R(t,t_1) \Sigma_\alpha^<(t_1,t') \Big\} .
\end{equation}
In Wigner representation the expression for the current becomes
\begin{equation}
J_\alpha(T) = \int_{-\infty}^{+\infty} d\omega \widetilde{C}_\alpha(T,\omega),
\end{equation}
where $\widetilde{C}_\alpha(T,\omega)$ is the Wigner transformation of the two-time function $C_\alpha(t,t') $:
\begin{equation}
\widetilde{C}_\alpha(T,\omega) =2 \text{Re} \Big\{ e^{\frac{1}{2i} ( \partial^{\cal G}_T \partial^\Sigma_\omega- \partial^{\cal G}_\omega \partial^\Sigma_T)} \Big( \widetilde{\cal G}^<\widetilde{\Sigma}_\alpha^A+ \widetilde{\cal G}^R \widetilde{\Sigma}_\alpha^< \Big) \Big\} .
\end{equation}
The above equation is altered by taking a second order gradient expansion for the exponential derivatives and expanding the Green's function up to the second order (note that we use the expansion for the lesser Green's function 
$ {\cal G}^< = G^<+ \widetilde{\cal G}^<_{(1)} + \widetilde{\cal G}^<_{(2)} $ similar to (\ref{expansion}) but the  particular form of the terms in this expansion is not required for our final expression). This allows us to break this equation for the current  based on order to get
\begin{equation}
J^{(0)}_{\alpha}(\mathbf q) = \frac{1}{\pi}\int^{\infty}_{-\infty} d \omega \text{Re} \Big\{ G^<\widetilde{\Sigma}_{\alpha}^{A} + G^R\widetilde{\Sigma}_{\alpha}^{<} \Big\},
\end{equation}
\begin{multline}
J^{(1)}_{\alpha}(\mathbf q, \dot{ \mathbf q}) = \frac{1}{\pi}\int^{\infty}_{-\infty} d \omega \text{Re} \Big\{ \widetilde{\mathcal{G}}^{<}_{(1)} \widetilde{\Sigma}_{\alpha}^{A} + \widetilde{\mathcal{G}}^{R}_{(1)} \widetilde{\Sigma}_{\alpha}^{<} + \frac{1}{2i} \Big( \partial_{T} G^R\Big) \Big( \partial_{\omega} \widetilde{\Sigma}_{\alpha}^{<} \Big) \\ - \frac{1}{2i} \Big( \partial_{\omega} G^<\Big) \Big( \partial_{T} \widetilde{\Sigma}_{\alpha}^{A} \Big) - \frac{1}{2i} \Big( \partial_{\omega} G^R\Big) \Big( \partial_{T} \widetilde{\Sigma}_{\alpha}^{<} \Big) \Big\}
\end{multline}
and
\begin{multline}
J^{(2)}_{\alpha}(\mathbf q, \dot{ \mathbf q}^2, \ddot{ \mathbf q})  = \frac{1}{\pi}\int^{\infty}_{-\infty} d \omega \text{Re} \Big\{ \widetilde{\mathcal{G}}^{<}_{(2)} \widetilde{\Sigma}_{\alpha}^{A} +\widetilde{\mathcal{G}}^{R}_{(2)} \widetilde{\Sigma}_{\alpha}^{<} + \frac{1}{2i} \Big( \partial_{T} \widetilde{\mathcal{G}}^{R}_{(1)} \Big) \Big( \partial_{\omega} \widetilde{\Sigma}_{\alpha}^{<} \Big) \\ - \frac{1}{2i} \Big( \partial_{\omega}\widetilde{\mathcal{G}}^{<}_{(1)} \Big) \Big( \partial_{T} \widetilde{\Sigma}_{\alpha}^{A} \Big) - \frac{1}{2i} \Big( \partial_{\omega} \widetilde{\mathcal{G}}^{R}_{(1)} \Big) \Big( \partial_{T} \widetilde{\Sigma}_{\alpha}^{<} \Big) - \frac{1}{8} \Big( \partial^{2}_{T} G^R\Big) \Big( \partial^{2}_{\omega} \widetilde{\Sigma}_{\alpha}^{<} \Big)  \\ + \frac{1}{4} \Big( \partial_{T \omega} G^R\Big) \Big( \partial_{\omega T} \widetilde{\Sigma}_{\alpha}^{<} \Big) - \frac{1}{8} \Big( \partial^{2}_{\omega} G^<\Big) \Big( \partial^{2}_{T} \widetilde{\Sigma}_{\alpha}^{A} \Big) - \frac{1}{8} \Big( \partial^{2}_{\omega} G^R\Big) \Big( \partial^{2}_{T} \widetilde{\Sigma}_{\alpha}^{<} \Big) \Big\}.
\end{multline}
It is useful to alter the form of the current equations using the identities
\begin{equation}
\int^{\infty}_{-\infty} d \omega \widetilde{A} ( \partial_{\omega} \widetilde{\Sigma}) = - \int^{\infty}_{-\infty} d \omega ( \partial_{\omega} \widetilde{A} ) \widetilde{\Sigma}
\end{equation}
and 
\begin{equation}
\int^{\infty}_{-\infty} d \omega \widetilde{A} ( \partial^{2}_{\omega} \widetilde{\Sigma} ) = \int^{\infty}_{-\infty} d \omega (\partial^{2}_{\omega} \widetilde{A} ) \widetilde{\Sigma},
\end{equation}
which apply for arbitrary Green's functions $\widetilde{A}$ and self energy $\widetilde{\Sigma}$ quantities and are a consequence of the fact that the Green's functions vanish as $| t - t^{\prime} | \rightarrow \pm \infty$. This allows us to express the equations for current as
\begin{equation}
J^{(0)}_{\alpha}(\mathbf q)  = \frac{1}{\pi}\int^{\infty}_{-\infty} d \omega \text{Re} \Big\{ G^<\widetilde{\Sigma}_{\alpha}^{A} + G^R\widetilde{\Sigma}_{\alpha}^{<} \Big\},
\end{equation}
\begin{multline}
J^{(1)}_{\alpha}(\mathbf q, \dot{ \mathbf q}) = \frac{1}{\pi}\int^{\infty}_{-\infty} d \omega \text{Re} \Big\{ \widetilde{\mathcal{G}}^{<}_{(1)} \widetilde{\Sigma}_{\alpha}^{A} + \widetilde{\mathcal{G}}^{R}_{(1)} \widetilde{\Sigma}_{\alpha}^{<} - \frac{1}{2i} \Big( \partial_{T \omega} G^R\Big) \widetilde{\Sigma}_{\alpha}^{<}   - \frac{1}{2i} \Big( \partial_{\omega} G^R\Big) \Big( \partial_{T} \widetilde{\Sigma}_{\alpha}^{<} \Big) \Big\}
\end{multline}
and
\begin{multline}
J^{(2)}_{\alpha}(\mathbf q, \dot{ \mathbf q}^2, \ddot{ \mathbf q})  = \frac{1}{\pi}\int^{\infty}_{-\infty} d \omega \text{Re} \Big\{ \widetilde{\mathcal{G}}^{<}_{(2)} \widetilde{\Sigma}_{\alpha}^{A} + \Big[ \widetilde{\mathcal{G}}^{R}_{(2)} - \frac{1}{8} \Big( \partial^{2}_{T \omega} G^R\Big) \Big] \widetilde{\Sigma}_{\alpha}^{<} \\ + \frac{1}{4} \Big( \partial_{T} \partial^{2}_{ \omega} G^R\Big) \Big( \partial_{T} \widetilde{\Sigma}_{\alpha}^{<} \Big) - \frac{1}{8} \Big( \partial^{2}_{\omega} G^R\Big) \Big( \partial^{2}_{T} \widetilde{\Sigma}_{\alpha}^{<} \Big) \Big\}.
\end{multline}
We now consider the net second order non-adiabatic current as 
\begin{equation}
J^{(2)}(\mathbf q, \dot{ \mathbf q}^2, \ddot{ \mathbf q}) = y(\mathbf q) J_{L}^{(2)}(\mathbf q, \dot{\mathbf q}, \ddot{\mathbf q}) - (1 - y(\mathbf q)) J_{R}^{(2)}(\mathbf q, \dot{\mathbf q}, \ddot{\mathbf q}), 
\end{equation}
where $y(\mathbf q)$ is an arbitrary function of $\mathbf q$. Function  $y(\mathbf q)$ can be chosen such that the final expression for the current has a particularly simple form.
Substituting in explicit expressions for $J_{L}^{(2)}(\mathbf q, \dot{ \mathbf q}^2, \ddot{ \mathbf q})  $ and $J_{R}^{(2)}(\mathbf q, \dot{ \mathbf q}^2, \ddot{ \mathbf q})  $ we find that $J^{(2)}(\mathbf q, \dot{ \mathbf q}^2, \ddot{ \mathbf q}) $ becomes 
\begin{multline}
J^{(2)}(\mathbf q, \dot{ \mathbf q}^2, \ddot{ \mathbf q})   = \frac{1}{\pi} \int^{\infty}_{-\infty}  d \omega \text{Re} \Big\{ \widetilde{G}^{<}_{(2)} \Big( y(\mathbf q) \Sigma^{A}_{L}- (1-y(\mathbf q)) \Sigma^{A}_{R} \Big) + \Big[ \widetilde{G}^{R}_{(2)} - \frac{1}{8} \partial^{2}_{T \omega}G^R\Big] \\ \Big( y(\mathbf q) \widetilde{\Sigma}^{<}_{L}- ( 1 - y(\mathbf q) ) \widetilde{\Sigma}^{<}_{R} \Big) + \frac{1}{4} \partial_{T} \partial^{2}_{\omega}G^R\Big( y(\mathbf q) \partial_{T} \widetilde{\Sigma}^{<}_{L} - ( 1 - y(\mathbf q) ) \partial_{T} \widetilde{\Sigma}^{<}_{R} \Big) \\ - \frac{1}{8} \partial^{2}_{\omega}G^R\Big( y(\mathbf q) \partial^{2}_{T} \widetilde{\Sigma}^{<}_{L} - ( 1 - y(\mathbf q) ) \partial^{2}_{T} \widetilde{\Sigma}^{<}_{R} \Big) \Big\}. 
\label{eq1}
\end{multline}
We now choose $y(\mathbf q)$ such that the lesser Green's function term disappears. This is done by solving for $y(\mathbf q)$ given
\begin{equation}
y(\mathbf q) \widetilde \Sigma^{A}_{L} - ( 1 - y(\mathbf q) ) \widetilde \Sigma^{A}_{R} = 0.
\end{equation}
This can be easily solved to give 
\begin{equation}
y(\mathbf q) = \frac{\widetilde \Sigma^{A}_{R}}{\widetilde \Sigma^{A}_{L}+ \widetilde \Sigma^{A}_{R}} = \frac{\gamma_{R} (1+ \lambda_R q_R)^2}{\gamma_{L} (1+ \lambda_L q_L)^2+ \gamma_{R} (1+ \lambda_R q_R)^2} = \frac{\Gamma_R}{\Gamma_L + \Gamma_R}.
\end{equation}
By making substitutions for $y(\mathbf q)$, we find that (\ref{eq1}) becomes 
\begin{multline}
J^{(2)}(\mathbf q, \dot{ \mathbf q}^2, \ddot{ \mathbf q})  = \frac{1}{\pi} \int^{+\infty}_{-\infty} d \omega \text{Re} \Big\{ \Big[ \widetilde{G}^{R}_{(2)} - \frac{1}{8} \partial^{2}_{T \omega}G^R\Big] \frac{\Gamma_{R} \widetilde{\Sigma}^{<}_{L} - \Gamma_{L} \widetilde{\Sigma}^{<}_{R}}{\Gamma_{L} + \Gamma_{R}} \\ + \frac{1}{4} \partial_{T} \partial^{2}_{\omega}G^R\frac{\Gamma_{R} \partial_{T} \widetilde{\Sigma}^{<}_{L} - \Gamma_{L} \partial_{T} \widetilde{\Sigma}^{<}_{R}}{\Gamma_{L} + \Gamma_{R}} - \frac{1}{8} \partial^{2}_{\omega}G^R\frac{\Gamma_{R} \partial^{2}_{T} \widetilde{\Sigma}^{<}_{L} - \Gamma_{L} \partial^{2}_{T} \widetilde{\Sigma}^{<}_{R}}{\Gamma_{L} + \Gamma_{R}} \Big\}. 
\end{multline}
We now make substitutions for the self-energy terms where we use the form 
\begin{equation}
\widetilde \Sigma^{A}_{\alpha} =  \frac{i}{2}  \Gamma_{\alpha}
\end{equation}
and 
\begin{equation}
\widetilde \Sigma^{<}_{\alpha} = i \Gamma_{\alpha} f_{\alpha},
\end{equation}
as was first defined in (\ref{advancedS}) and (\ref{lesserS}) respectively. We find that 
\begin{multline}
J^{(2)}(\mathbf q, \dot{ \mathbf q}^2, \ddot{ \mathbf q})  = \frac{1}{\pi \Gamma} \int^{+\infty}_{-\infty} d \omega \text{Re} \Big\{ i \Gamma_{L} \Gamma_{R} \Big[ \widetilde{G}^{R}_{(2)} - \frac{1}{8} \partial^{2}_{T \omega}G^R \Big] \Big( f_{L} - f_{R} \Big) \\ + \frac{i}{4} \partial_{T} \partial^{2}_{\omega} G^R \Big( \dot{\Gamma}_{L} \Gamma_{R} f_{L} - \Gamma_{L} \dot{\Gamma}_{R} f_{R} \Big) - \frac{i}{8} \partial^{2}_{\omega}G^R\Big( \ddot{\Gamma}_{L} \Gamma_{R} f_{L} - \Gamma_{L} \ddot{\Gamma}_{R} f_{R} \Big) \Big\}. 
\end{multline}
In the equation above we have used the quantity 
\begin{equation}
\Gamma = \Gamma_{L} + \Gamma_{R}.
\end{equation}
We now take a detour and search for explicit forms for all derivatives of $\Gamma_{\alpha}$ quantities. Reminding the reader that $\Gamma_{\alpha}$ takes the explicit form (see (\ref{115})) 
\begin{equation}
\Gamma_{\alpha} = \gamma_{\alpha}(1 + \lambda_{\alpha} q_\alpha)^2,
\end{equation}
then we can show that its central time derivatives are given by
\begin{equation}
\dot{\Gamma}_{\alpha} = 2 \gamma_{\alpha} \lambda_{\alpha} \dot{q}_{\alpha} (1 + \lambda_{\alpha} q_\alpha)
\label{171}
\end{equation}
and 
\begin{equation}
\ddot{\Gamma}_{\alpha} = 2 \gamma_{\alpha} \lambda_{\alpha}^2 \dot{q}_{\alpha}^2 + 2 \gamma_{\alpha} \lambda_{\alpha} \ddot{q}_{\alpha} (1 + \lambda_{\alpha} q_\alpha).
\label{172}
\end{equation} 
Derivatives of $\Gamma_{\alpha}$ allow one to compute derivatives of the retarded self-energy component (these will be important later on in the derivation) which we find are given by
\begin{equation}
\partial_T \widetilde{\Sigma}^{R}_{\alpha} = - i \gamma_{\alpha} \lambda_{\alpha} \dot{q}_{\alpha} (1 + \lambda_{\alpha} q_\alpha)
\end{equation}
and 
\begin{equation}
\partial^2_T \widetilde{\Sigma}^{R}_{\alpha} = - i \gamma_{\alpha} \lambda_{\alpha}^2 \dot{q}_{\alpha}^2 - i \gamma_{\alpha} \lambda_{\alpha} \ddot{q}_{\alpha} (1 + \lambda_{\alpha} q_\alpha).
\end{equation} 

Making substitutions for (\ref{171}) and (\ref{172}) (neglecting acceleration terms since they will disappear once averaged over time), we find that
\begin{multline}
J^{(2)}(\mathbf q, \dot{ \mathbf q}^2)  = \frac{\gamma_{L} \gamma_{R}}{\pi \Gamma} \int^{+\infty}_{-\infty} d \omega \text{Re} \Big\{ i (1 + \lambda_L q_L)^2 (1 + \lambda_R q_R)^2 \Big[ \widetilde{G}^{R}_{(2)} - \frac{1}{8} \partial^{2}_{T \omega}G^R\Big] \Big( f_{L} - f_{R} \Big) \\ + \frac{i}{2} (1 + \lambda_L q_L) (1 + \lambda_R q_R) \partial_{T} \partial^{2}_{\omega}G^R\Big( \dot{q}_{L} \lambda_L (1 + \lambda_R q_R) f_{L} -  \dot{q}_{R} \lambda_R (1 + \lambda_L q_L) f_{R} \Big) \\ - \frac{i}{4} \partial^{2}_{\omega}G^R\Big( \dot{q}_{L}^2 \lambda_{L}^{2} (1 + \lambda_R q_R)^2 f_{L} - \dot{q}_{R}^2 \lambda_{R}^{2} (1 + \lambda_L q_L)^2 f_{R} \Big) \Big\}. 
\end{multline}
In order to simplify the presentation of the  expressions we introduce the quantity
\begin{equation}
\Lambda_\alpha = 1 + \lambda_\alpha q_\alpha.
\end{equation}
We now substitute in the explicit forms of $\widetilde{G}^{R}_{(2)}$ and all Green's function derivatives to find
\begin{multline}
J^{(2)}(\mathbf q, \dot{ \mathbf q}^2) = \frac{\gamma_{L} \gamma_{R}}{\pi \Gamma} \int^{+\infty}_{-\infty} d \omega \text{Re} \Big\{- i \Lambda_L^2 \Lambda_R^2 \Big[ \Big(G^R\Big)^{4} \partial_{T}^2 \Sigma^{R} + 3 \Big(G^R\Big)^{5} \Big( \partial_{T} \Sigma^{R} \Big)^2 \Big] \Big( f_{L} - f_{R} \Big) \\ + 3 i \Lambda_L \Lambda_R \Big(G^R\Big)^{4} \partial_{T} \Sigma^R \Big( \dot{q}_{L} \lambda_L \Lambda_R f_{L} -  \dot{q}_{R} \lambda_R \Lambda_L f_{R} \Big) \\ - \frac{i}{2} \Big(G^R\Big)^{3} \Big( \dot{q}_{L}^2 \lambda_{L}^{2} \Lambda_R^2 f_{L} - \dot{q}_{R}^2 \lambda_{R}^{2} \Lambda_L^2 f_{R} \Big) \Big\}. 
\end{multline} 
Making a substitution for the self-energy components yields (with some rearrangement)
\begin{multline}
J^{(2)}(\mathbf q, \dot{ \mathbf q}^2)  = \frac{\Gamma_{L} \Gamma_{R}}{\pi \Gamma} \int^{+\infty}_{-\infty} d \omega \text{Re} \Big\{- \Big[ \Big(G^R\Big)^{4} \Big( \dot{q}^{2}_{L} \lambda_L^2 \gamma_L  + \dot{q}^{2}_{R} \lambda_R^2 \gamma_R \Big) \\ - 3 i \Big(G^R\Big)^{5} \Big( \dot{q}_{L} \lambda_L \gamma_L \Lambda_L  + \dot{q}_{R} \lambda_R \gamma_R \Lambda_R \Big)^2 \Big] \Big( f_{L} - f_{R} \Big) \\ + \frac{3}{\Lambda_L \Lambda_R} \Big(G^R\Big)^{4} \Big( \dot{q}_{L} \lambda_L \gamma_L \Lambda_L  + \dot{q}_{R} \lambda_R \gamma_R \Lambda_R \Big) \Big( \dot{q}_{L} \lambda_L \Lambda_R f_{L} -  \dot{q}_{R} \lambda_R \Lambda_L f_{R} \Big) \\ - \frac{i}{2 \Lambda_L^2 \Lambda_R^2} \Big(G^R\Big)^{3} \Big( \dot{q}_{L}^2 \lambda_{L}^{2} \Lambda_R^2 f_{L} - \dot{q}_{R}^2 \lambda_{R}^{2} \Lambda_L^2 f_{R} \Big) \Big\}. 
\label{j2}
\end{multline} 
The second order (in nuclear velocities) corrections to the electric current (\ref{j2}) is one of the main results of the paper.
The total electric current combines 
the zeroth order adiabatic electric current, which depends only on the instantaneous nuclear geometry,
and the second order velocity-dependent non-adiabatic term (\ref{j2}) to give
\begin{equation}
\label{j0}
J^{(0)}(\mathbf q) =- \frac{\Gamma_{L} \Gamma_{R}}{\pi \Gamma} \int^{+\infty} _{-\infty} d \omega \text{Im} \Big\{ G^R \Big\} \Big( f_{L} - f_{R} \Big).
\end{equation}

\section{Results}

The proposed theory is illustrated in the single molecular orbital and single nuclear degree of freedom case. We choose parameters $q=q_L=-q_R$ which means that if the left bond stretches then the right bond contracts by the same amount and vice versa; we also assume that $\lambda = \lambda_L = \lambda_R$. 
In this limit, we find that by taking (\ref{j0}, \ref{j2}) and removing the terms from the second-order non-adiabatic corrections that violate the current conservation (the details are discussed in the Appendix), we can compute the total current
\begin{equation}
J(q, \overline{\dot{  q}^2})  =  J^{(0)}( q) +  J^{(2)}( q,\overline{\dot{ q}^2}) = \int_{- \infty}^{+ \infty} d\omega {\cal T} (q, \overline{\dot{ q}^2}, \omega) \Big( f_L - f_R \Big),
\label{jtot}
\end{equation}
with the function ${\cal T} (q, \overline{\dot{ q}^2}, \omega) $ defined according to
\begin{multline}
\label{ttty}
{\cal T} ( q,\overline{\dot{  q}^2}, \omega)  = - 2_s \frac{\Gamma_{L} \Gamma_{R}}{\pi \Gamma} \Big[ \text{Im} \Big\{ G^R \Big\} + \overline{\dot q^2} \lambda^2 \Big(\gamma_L + \gamma_R\Big) \text{Re} \Big\{ \Big( G^R \Big)^{4} \Big\}  \\  + 3 \overline{\dot q^2} \lambda^2 \Big( ( \gamma_L - \gamma_R ) + \lambda q ( \gamma_L + \gamma_R ) \Big)^2 \text{Im} \Big\{ \Big(G^R\Big)^{5} \Big\} - 3 q \frac{\overline{\dot q^2} \lambda^3}{\Lambda_L \Lambda_R} \Big( ( \gamma_L - \gamma_R ) \\ +  \lambda q ( \gamma_L + \gamma_R )  \Big) \text{Re} \Big\{ \Big(G^R \Big)^{4} \Big\} -  \frac{\overline{\dot q^2} \lambda^2}{2\Lambda_L \Lambda_R} \text{Im} \Big\{ \Big(G^R\Big)^{3} \Big\} \Big].
\end{multline}
Here we have used the notation $2_s$ to denote the electronic spin degeneracy of the system.
A small word on the interpretation of equations (\ref{ttty}) and (\ref{jtot}). Note that despite its similar appearance to the Landauer formula, ${\cal T} ( q, \overline{\dot{  q}^2},\omega)$ should not be viewed as the transmission probability for an electron with energy $\omega$. The inelastic effects are present in our model (interaction between electronic and nuclear degrees of freedom) and therefore, generally, there is no connection between ${\cal T} (q, \overline{\dot{  q}^2},\omega)$ given by (\ref{ttty}) and the probability for an electron with energy $\omega$ to be transferred across the molecular bridge.

Let us first  take the molecular geometry at the equilibrium configuration $q=0$  and $\gamma_L=\gamma_R$. The physical meaning of expression  (\ref{ttty}) becomes particularly apparent in this case. Computing  (\ref{ttty}) at $\omega=0$ gives the expression for the molecular conductance
\begin{equation}
G =\frac{2_s}{2 \pi}  \frac{ (\gamma/2)^2}{\epsilon^2 + (\gamma/2)^2}  \Big( 1 - \overline{ \dot q^2} \lambda^2 \; \frac{ 7\epsilon^4  - 22 (\gamma/2)^2 \epsilon^2 + 3 (\gamma/2)^4}{2 [\epsilon^2 + (\gamma/2)^2]^3}  \Big),
\label{G}
\end{equation}
where we have used the quantity $\gamma$ defined by $\gamma = \gamma_L + \gamma_R.$ The first term in (\ref{G})  is the standard adiabatic expression for the conductance for the single resonant-level with the "frozen" geometry. The second term is proportional to the squared  nuclear velocity and describes the non-adiabatic correction to the molecular conductance. The non-adiabatic correction can either increase or decrease the molecular conductance depending on the level position $\epsilon$ relative to the Fermi energy (the Fermi energy is set to zero in our calculations). The sign of the non-adiabatic correction is determined by the biquadratic function $ 7\epsilon^4  - 22 (\gamma/2)^2 \epsilon^2 + 3(\gamma/2)^4 $, which has three positive (destructive contribution) and two negative (constructive contribution) regions as a function of $\epsilon$.

Through the use of (\ref{G}), we evaluate the molecular conductivity to investigate the adiabatic and non-adiabatic  contributions and Fig. 1 shows the result of calculations. We choose level broadening  $\gamma_L = \gamma_R = 0.5$ and average nuclear velocity $\overline{\dot q^2} = 0.1$, with these quantities being selected to reflect possible experimental values.  Calculations are conducted for a range of values of the coupling strength $\lambda$ with values $\lambda = 0,1,2$ being considered. Here $\lambda = 0$ will correspond to fully adiabatic transport and $\lambda= 1, 2$ will correspond to non-adiabatic transport.
The non-adiabatic motion on the molecule-electrode interface always suppresses the conductance in the resonant tunneling regime when the molecular orbital energy lies in the vicinity of the electrode Fermi energy. Once the molecular level is shifted away from the resonance, either above or below the Fermi energy, the non-adiabatic nuclear motion starts to play a constructive role by promoting the transport of electrons across the molecule. It is interesting to note that the molecular conductance becomes larger than $G_0= 2e^2/h $ in the strong coupling regime (given by $\lambda=2$)  indicating that the non-adiabatic nuclear motion opens an extra transport channel in this situation.  Then, far away from the resonance, the non-adiabatic corrections decrease the conductance but their effect is very marginal here.
\begin{figure}[t!]
\begin{center}
\includegraphics[width=1.0\columnwidth]{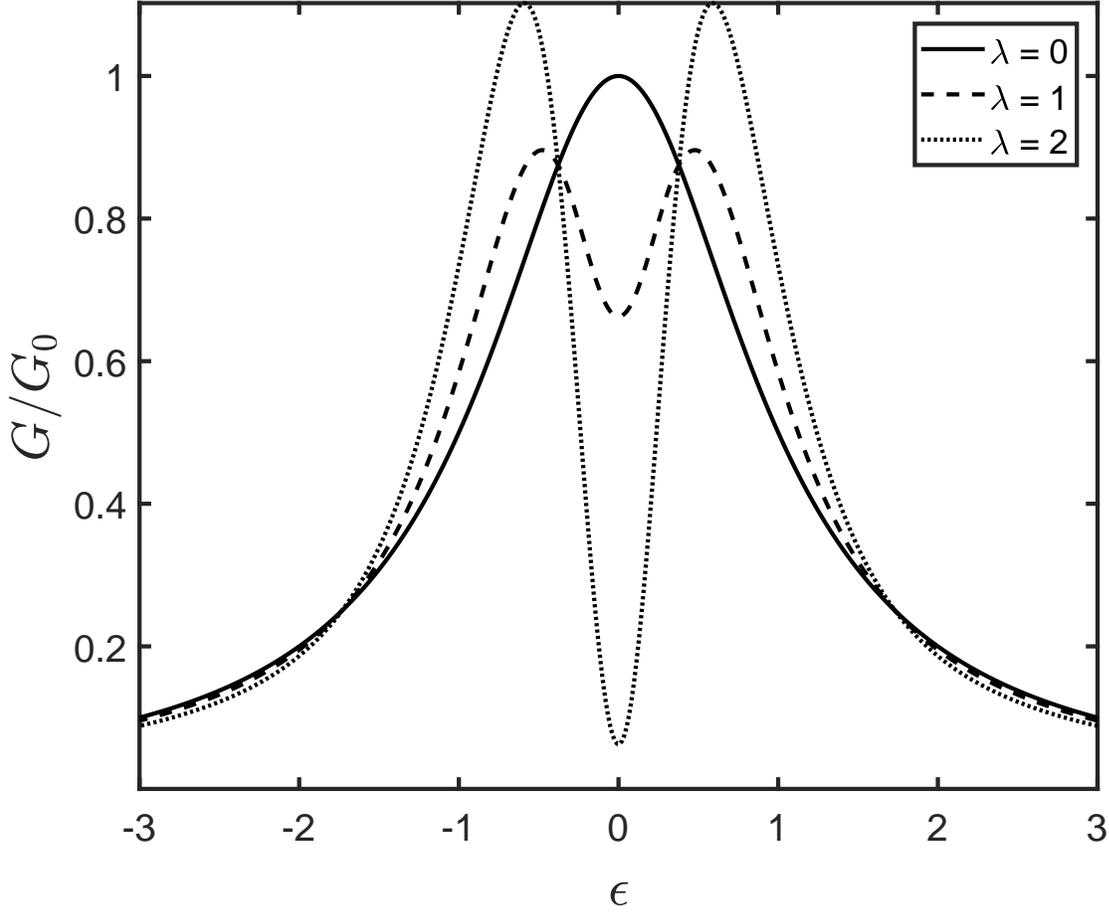}
\end{center}
	\caption{Conductance computed  for different electron-nuclei coupling strength $\lambda$. The values of the conductance  is given in terms of  $G_0= 2e^2/h $. 
	Parameters used in calculations: 
$\gamma_L=\gamma_R=0.5$, $\overline {\dot q_L^2}=\overline {\dot q_R^2}=0.1$, $\epsilon=0$.}
\label{figure1}
\end{figure}

Let us now consider the case of electron transport where the molecular junction interface is no longer confined to its equilibrium geometry. We do this by averaging the system over nuclear position and velocity according to the Boltzmann factor for a quadratic potential $V(q) = \frac{1}{2} k q^2$, where the spring constant $k$ describes the rigidity of the molecule-electrode linkage bonds. In Fig. 2 we plot the transmission coefficient for three values of the spring constant $k = 0.2, 2, 20$. 
\begin{figure}[t!]
\begin{center}
\includegraphics[width=1.0\columnwidth]{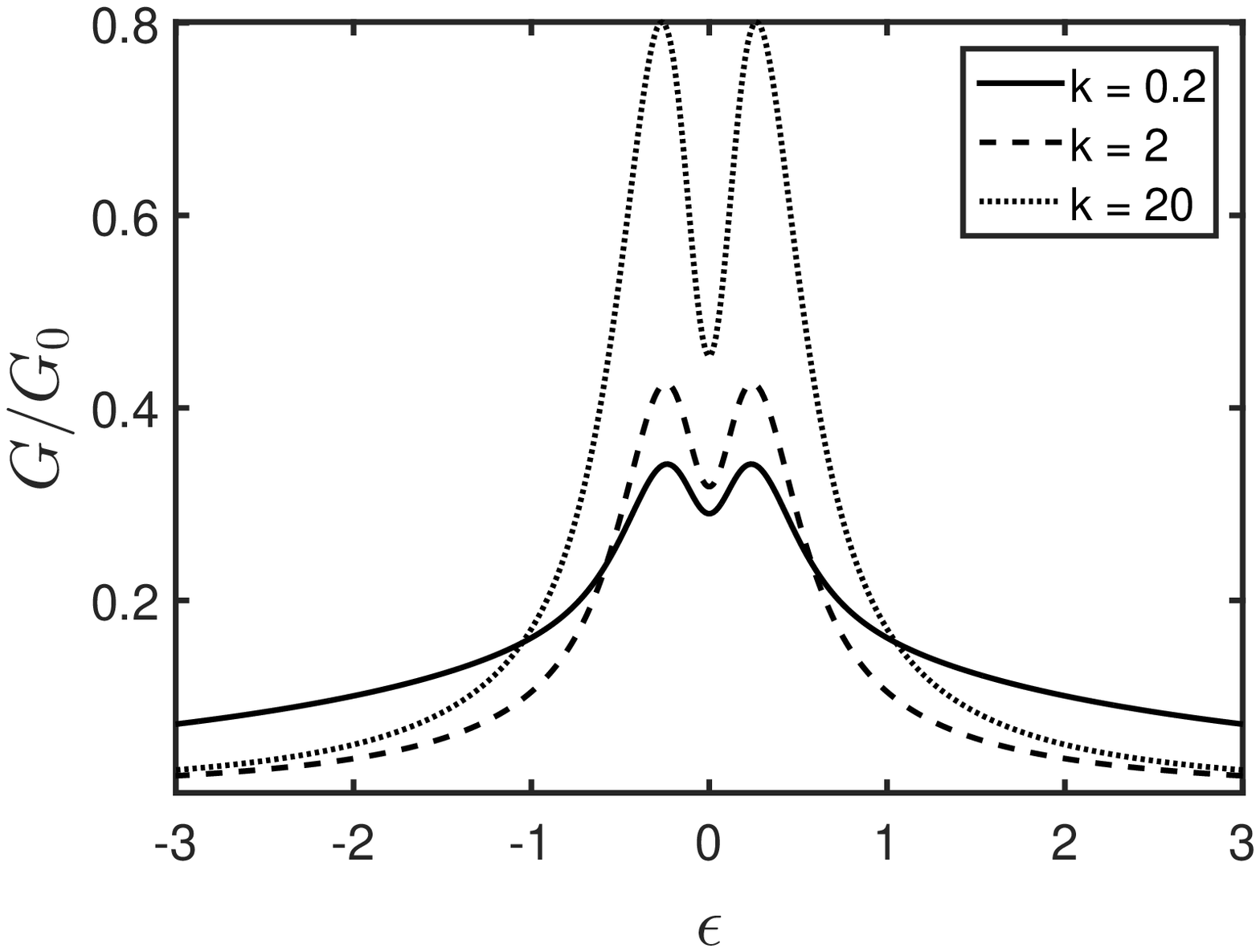}
\end{center}
	\caption{Conductance computed for different values of the electrode-molecular bond spring constant $k$. The values of the conductance  is given in terms of  $G_0= 2e^2/h $. Parameters used in calculations: 
$\gamma_L = \gamma_R=0.5$, $\overline {\dot q_L^2}=\overline {\dot q_R^2}=0.1$, $\epsilon=0$, $\lambda=1$.}
\label{figure2}
\end{figure}
We see in Fig. 2 that the non-adiabatic effects take on larger values for molecular junctions of increasing rigidity of molecule-electrode interfaces. Overall, the softer molecule-electrode bonds make the molecule less conductive unless the molecular orbital is shifted away from the electrode Fermi energy, where the effect is less pronounced and reversed.

 The current consensus in molecular electronics is that, in the  off-resonant situations, the typical signature of vibrational
modes is a small increase of the differential conductance. In resonance regimes, however, the nuclear motion manifests itself as a  small drop in the conductance.\cite{moletronics}
The results of our theory are in general qualitative agreement with these observations.

\section{Conclusion}
This paper has detailed a quantum transport theory that computes the electronic current while taking into account the non-adiabatic dynamical effects of motion of atoms on the molecule-electrode interfaces. Our approach makes use of the Keldysh non-equilibrium Green's functions technique where the equations of motion are mapped into the Wigner space such that fast and slow time scales are easily identifiable. The equations of motion are then solved in the limit that interfacial nuclear dynamics are slow where, as a result, a systematic perturbative expansion is developed about the small parameter to compute the adiabatic molecular Green's functions with first and second order non-adiabatic corrections. These components are used to compute the electric current as a function of molecular geometry, velocities and accelerations of nuclei contained in the molecule-lead interface. Our equations allow for the calculations of electronic transport characteristics of molecular junctions where we do not need to assume that the molecular deformation about the equilibrium geometry is small or harmonic, neither do we need to assume that coupling between the nuclear and electronic degrees of freedom is small.

The proposed theory is applied to a simple transport model with a single molecular and a single classical degree of freedom. We find that the motion of nuclei in the molecule-lead interface  result into molecular junctions which are less or more transmissive for electron tunneling  depending on the position of the molecular orbital energy relative to the electrode Fermi energy. We find that the non-adiabatic effects generally decreases the molecular conductance if the molecule orbital is aligned with the electrode Fermi energy, but 
play the constructive role by opening extra transport channels and increasing the conductance once we shift the energy level away from the resonance.


\appendix
\section{Conserving and non-conserving parts of the electric current}
From the main body of the paper we know that 
\begin{equation}
\label{jbj}
J(\mathbf q) =- \frac{\Gamma_{L} \Gamma_{R}}{\pi \Gamma} \int^{+\infty}_{-\infty} d \omega \Big( f_{L} - f_{R} \Big) \text{Im} \Big\{ G^R \Big\}
\end{equation}
and
\begin{multline}
\label{xx}
J^{(2)}(\mathbf q, \dot{ \mathbf q}^2)  = - \frac{\Gamma_{L} \Gamma_{R} }{\pi \Gamma} \int^{+\infty}_{-\infty} d \omega \Big[ \Big( \dot{q}^{2}_{L} \lambda_L^2 \gamma_L  + \dot{q}^{2}_{R} \lambda_R^2 \gamma_R \Big) \Big( f_{L} - f_{R} \Big) \text{Re} \Big\{ \Big(G^R\Big)^{4} \Big\} \\ + 3 \Big( \dot{q}_{L} \lambda_L \gamma_L \Lambda_L  + \dot{q}_{R} \lambda_R \gamma_R \Lambda_R \Big)^2 \Big( f_{L} - f_{R} \Big)  \text{Im} \Big\{ \Big(G^R\Big)^{5} \Big\}  \\ + \frac{3}{\Lambda_L \Lambda_R} \Big( \dot{q}_{L} \lambda_L \gamma_L \Lambda_L  + \dot{q}_{R} \lambda_R \gamma_R \Lambda_R \Big) \Big( \dot{q}_{L} \lambda_L \Lambda_R f_{L} -  \dot{q}_{R} \lambda_R \Lambda_L f_{R} \Big) \text{Re} \Big\{ \Big(G^R\Big)^{4} \Big\} \\ - \frac{1}{2\Lambda^2_L \Lambda^2_R} \Big( \dot{q}_{L}^2 \lambda_{L}^{2} \Lambda_R^2 f_{L} - \dot{q}_{R}^2 \lambda_{R}^{2} \Lambda_L^2 f_{R} \Big) \text{Im} \Big\{ \Big(G^R\Big)^{3} \Big\} \Big], 
\end{multline}
where we have neglected the velocity dependent term in (\ref{xx}). It follows that one can specify the total current from the left lead according to 
\begin{multline}
J(\mathbf q, \dot{ \mathbf q}^2) =  - \frac{\Gamma_{L} \Gamma_{R}}{\pi \Gamma} \int^{+\infty}_{-\infty} d \omega \Big[ \Big( f_{L} - f_{R} \Big) \text{Im} \Big\{ G^R \Big\} + \Big( \dot{q}^{2}_{L} \lambda_L^2 \gamma_L  + \dot{q}^{2}_{R} \lambda_R^2 \gamma_R \Big) \\ \times \Big( f_{L} - f_{R} \Big) \text{Re} \Big\{ \Big(G^R\Big)^{4} \Big\} + 3 \Big( \dot{q}_{L} \lambda_L \gamma_L \Lambda_L + \dot{q}_{R} \lambda_R \gamma_R \Lambda_R \Big)^2 \Big( f_{L} - f_{R} \Big)  \text{Im} \Big\{ \Big(G^R\Big)^{5} \Big\}  \\ + \frac{3}{\Lambda_L \Lambda_R} \Big( \dot{q}_{L} \lambda_L \gamma_L \Lambda_L + \dot{q}_{R} \lambda_R \gamma_R \Lambda_R \Big) \Big( \dot{q}_{L} \lambda_L \Lambda_R f_{L} -  \dot{q}_{R} \lambda_R \Lambda_L f_{R} \Big) \\ \times \text{Re} \Big\{ \Big(G^R\Big)^{4} \Big\}  - \frac{1}{2 \Lambda^2_L \Lambda^2_R} \Big( \dot{q}_{L}^2 \lambda_{L}^{2} \Lambda_R^2 f_{L} - \dot{q}_{R}^2 \lambda_{R}^{2} \Lambda_L^2 f_{R} \Big) \text{Im} \Big\{ \Big(G^R\Big)^{3} \Big\} \Big].
\end{multline}
We now assume that $\lambda = \lambda_L = \lambda_R$ and $q = q_L = - q_R$. It follows that $\dot{q} = \dot{q}_L = - \dot{q}_R$ and that the square of the velocities is equated by $\dot{q}^2 = \dot{q}_L^2 = \dot{q}_R^2$. It is also relevant to note that under this assumption $\Lambda_L = 1 + \lambda q$ and $\Lambda_R = 1 - \lambda q$. As a result we can write 
\begin{multline}
J(q, \dot{q}^2) =  - \frac{\Gamma_{L} \Gamma_{R}}{\pi \Gamma} \int^{+\infty}_{-\infty} d \omega \Big[ \Big( f_{L} - f_{R} \Big) \text{Im} \Big\{ G^R \Big\} + \dot{q}^{2} \lambda^2 \Big( \gamma_L + \gamma_R \Big) \Big( f_{L} - f_{R} \Big) \\ \times \text{Re} \Big\{ \Big(G^R\Big)^{4} \Big\} + 3 \dot{q}^{2} \lambda^2 \Big( \gamma_L \Lambda_L - \gamma_R \Lambda_R \Big)^2 \Big( f_{L} - f_{R} \Big)  \text{Im} \Big\{ \Big(G^R\Big)^{5} \Big\}  \\ + 3 \frac{\dot{q}^{2} \lambda^2}{\Lambda_L \Lambda_R} \Big(  \gamma_L \Lambda_L  - \gamma_R \Lambda_R \Big) \Big( \Lambda_R f_{L} +  \Lambda_L f_{R} \Big) \text{Re} \Big\{ \Big(G^R\Big)^{4} \Big\} \\ -  \frac{\dot{q}^{2} \lambda^2}{2 \Lambda^2_L \Lambda^2_R}  \Big( \Lambda_R^2 f_{L} -  \Lambda_L^2 f_{R} \Big) \text{Im} \Big\{ \Big(G^R\Big)^{3} \Big\} \Big].
\end{multline}
In the equation above we notice that the first two terms preserve current conservation while the second two terms do not due to the presence of the $\Lambda_\alpha$ prefactors of the Fermi-Dirac distributions. By substituting for their explicit expressions and rearranging it is found that
\begin{multline}
J(q, \dot{q}^2) =  - \frac{\Gamma_{L} \Gamma_{R} }{\pi \gamma} \int^{+\infty}_{-\infty} d \omega \Big[ \Big( f_{L} - f_{R} \Big) \text{Im} \Big\{ G^R \Big\} + \dot{q}^{2} \lambda^2 \Big( \gamma_L + \gamma_R \Big) \Big( f_{L} - f_{R} \Big) \\ \times \text{Re} \Big\{ \Big(G^R\Big)^{4} \Big\} + 3 \dot{q}^{2} \lambda^2 \Big( \gamma_L \Lambda_L - \gamma_R \Lambda_R \Big)^2 \Big( f_{L} - f_{R} \Big)  \text{Im} \Big\{ \Big(G^R\Big)^{5} \Big\}  \\ + 3 \frac{\dot{q}^{2} \lambda^2}{\Lambda_L \Lambda_R} \Big(  \gamma_L \Lambda_L  - \gamma_R \Lambda_R \Big) \Big( ( f_{L} + f_{R} ) - \lambda q \big( f_{L} -  f_{R} \big) \Big) \text{Re} \Big\{ \Big(G^R\Big)^{4} \Big\} \\ -  \frac{\dot{q}^{2} \lambda^2}{2 \Lambda^2_L \Lambda^2_R} \Big( \Lambda_L \Lambda_R (f_{L} -  f_{R}) - 2 \lambda q (f_{L} +  f_{R})  \Big) \text{Im} \Big\{ \Big(G^R\Big)^{3} \Big\} \Big].
\end{multline}
This allows us to split the current equation given above into conserving and non-conserving components which we denote by $J_C(q, \dot{q}^2)$ and $J_{NC}(q, \dot{q}^2)$ according to 
\begin{equation}
J(q, \dot{q}^2) = J_{C}(q, \dot{q}^2) + J_{NC}(q, \dot{q}^2).
\end{equation}
We find that 
\begin{multline}
J_C(q, \dot{q}^2) =  - \frac{\Gamma_{L} \Gamma_{R} }{\pi \Gamma} \int^{+\infty}_{-\infty} d \omega \Big[ \text{Im} \Big\{ G^R \Big\} + \dot{q}^{2} \lambda^2 \Big( \gamma_L + \gamma_R \Big) \text{Re} \Big\{ \Big(G^R\Big)^{4} \Big\} \\  + 3 \dot{q}^{2} \lambda^2 \Big( \gamma_L \Lambda_L - \gamma_R \Lambda_R \Big)^2 \text{Im} \Big\{ \Big(G^R\Big)^{5} \Big\}  - 3 \lambda q \frac{\dot{q}^{2} \lambda^2}{\Lambda_L \Lambda_R} \Big(  \gamma_L \Lambda_L  - \gamma_R \Lambda_R \Big) \text{Re} \Big\{ \Big(G^R\Big)^{4} \Big\} \\ -  \frac{\dot{q}^{2} \lambda^2}{2 \Lambda_L \Lambda_R} \text{Im} \Big\{ \Big(G^R\Big)^{3} \Big\} \Big] \Big( f_{L} -  f_{R} \Big)
\label{C}
\end{multline}
and
\begin{multline}
J_{NC}(q, \dot{q}^2) = - \frac{\Gamma_{L} \Gamma_{R}}{\pi \Gamma} \int^{+\infty}_{-\infty} d \omega \Big[ - 3 \frac{\dot{q}^{2} \lambda^2}{\Lambda_L \Lambda_R} \Big( \gamma_L \Lambda_L  - \gamma_R \Lambda_R \Big) \text{Re} \Big\{ \Big(G^R\Big)^{4} \Big\} \\ + \lambda q \frac{\dot{q}^{2} \lambda^2}{\Lambda^2_L \Lambda^2_R} \text{Im} \Big\{ \Big(G^R\Big)^{3} \Big\} \Big] \Big( f_{L} +  f_{R} \Big).
\label{NC}
\end{multline}

\clearpage

%

\end{document}